\documentclass[a4paper,11pt]{article}
\usepackage{jinstpub} 
\usepackage{lineno}
\usepackage{makecell}

\title{Modeling Performance of Data Collection Systems for High-Energy Physics}

\author[a]{Wilkie Olin-Ammentorp\note{Corresponding author.},}
\author[a]{Xingfu Wu,}
\author[a,b]{and Andrew A. Chien}

\affiliation[a]{Argonne National Laboratory,\\
    9500 S. Cass St, Lemont, Illinois, USA
}
\affiliation[b]{University of Chicago,\\
    5801 S Ellis Ave, Chicago, Illinois, USA
}

\emailAdd{wolinammentorp@anl.gov}

\abstract{Exponential increases in scientific experimental data are outstripping the rate of progress in silicon technology. As a result, heterogeneous combinations of architectures and process or device technologies are increasingly important to meet the computing demands of future scientific experiments. However, the complexity of heterogeneous computing systems requires systematic modeling to understand performance. 

We present a model which addresses this need by framing key aspects of data collection pipelines and constraints, and combines them with the important vectors of technology that shape alternatives, computing metrics that allow complex alternatives to be compared. For instance, a data collection pipeline may be characterized by parameters such as sensor sampling rates, amount of data collected, and the overall relevancy of retrieved samples. Alternatives to this pipeline are enabled by hardware development vectors including advancing CMOS, GPUs, neuromorphic computing, and edge computing. By calculating metrics for each alternative such as overall F1 score, power, hardware cost, and energy expended per relevant sample, this model allows alternate data collection systems to be rigorously compared. 

To demonstrate this model's capability, we apply it to the CMS experiment (and planned HL-LHC upgrade) to evaluate and compare the application of novel technologies in the data acquisition system (DAQ). We demonstrate that improvements to early stages in the DAQ are highly beneficial, greatly reducing the resources required at later stages of processing (such as a 60\% power reduction) and increasing the amount of relevant data retrieved from the experiment per unit power (improving from 0.065 to 0.31 samples/$kJ$) However, we predict further advances will be required in order to meet overall power and cost constraints for the DAQ.}

\keywords{Data acquisition concepts, Analysis and statistical methods, Trigger concepts and systems (hardware and software)}

\begin{document}
\maketitle
\flushbottom

\section{Introduction}
Progress in many scientific disciplines is dependant upon the ability to collect and analyze increasingly large volumes of experimental data. Scientific experiments including X-ray microscopy, radio astronomy,  characterizing neutrinos, and exploring high-energy physics (HEP) will produce more data than in previous decades \cite{hammer_strategies_2021,jongerius_end--end_nodate,paul_laycock_for_the_dune_collaboration_dune_2021,das_overview_2022}. HEP in particularly is data-intensive, where detectors currently produce over 80 TB/s, with future increases planned \cite{cms_collaboration_phase-2_2021} This growth is a natural result of progress: refinement of extant knowledge requires higher precision, and discovery of new phenomena requires searching further and in greater detail than previously possible; achieving either goal is dependent on an increased ability to carry out experiments and process data. 

Historically, increased demand for data processing in scientific experiments was accompanied by multiple improvements in computing technologies, particularly in the areas of processing and communication. Processors became faster, memory became less expensive, and communication links improved in capacity and efficiency \cite{rupp_42_2018}. These advances enabled processing and transmitting greater amounts of data without a corresponding increase in power and space requirements. However, shifts in trends of microelectronic development - such as the end of Dennard scaling and the slow-down in Moore scaling - have reduced the rate of improvement in computing \cite{more_moore_team_international_nodate}. 

This has motivated new trends in computer architecture, where components and devices specialized at solving certain subsets of problems are becoming commonplace (such as graphics processors, network processors, neuromorphic systems, and more) \cite{hennessy_new_2019}. The availability of these components provides new opportunities for the designers of computing systems. However, evaluating the application of these components - where and how they should be integrated into computing systems to provide advantages over conventional approaches - represents a novel challenge. 

We provide an analytical, predictive model to evaluate the opportunities offered by novel hardware for computing systems. This "SystemFlow" model utilizes properties from individual hardware components to estimate the overall flow of information through a real-time computing system. Identifying these properties allows researchers from multiple disciplines, from devices to algorithms, to examine the impact their contributions can make towards improving system-level properties such as total power, effectiveness at retrieving relevant experimental data, number of computing devices required. This empowers researchers to systematically evaluate the role of emerging devices and architectures within large-scale computing systems. 

We demonstrate the application of this model to examine the performance of HEP data acquisition systems at the Compact Muon Solenoid (CMS) experiment at the Large Hadron Collider (LHC) \cite{cms_collaboration_cms_2008}, and quantify the benefits of multiple component-level changes. These include introducing novel data reduction processors, improving the skill of classifiers, the integration of parallelized algorithms, and varying system-level parameters.

\section{Background}
Computing empowers many areas of experimental science by automating the collection and analysis of data, providing feedback for control, real-time visualization, and more. Historically, a variety of computing technologies have been employed, but digital logic has come to dominate the field of computing. However, recent slow-downs in the scaling of digital logic have motivated alternative approaches. In the following sections, we detail these trends, a data-dependant scientific discipline, and how these fields interact to motivate the need for a systematic approach towards evaluating novel computing systems. 

\subsection{Computing Trends}
\label{sec:computing_trends}
Historically, increased demand for data processing in scientific experiments was accompanied by multiple improvements in computing technologies, particularly in the areas of processing and communication. Processors became faster, memory became less expensive, and communication links have improved in capacity and efficiency. These advances enabled processing and transmitting greater amounts of data without a corresponding increase in power and space requirements. 

\begin{figure}[ht]
\centering
\includegraphics[width=0.6\textwidth]{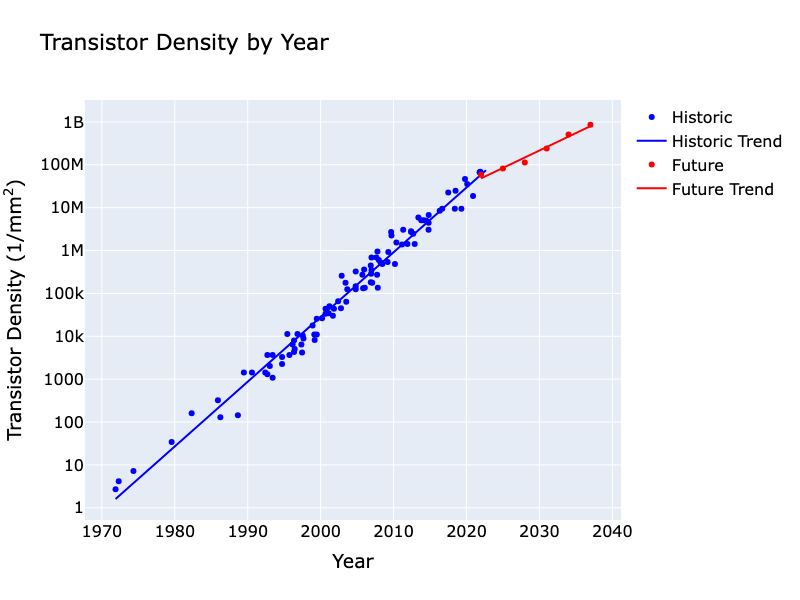}
\caption{Increases to the areal density of transistors which can be manufactured has increased exponentially over time, a phenomenon referred to as "Moore's Law." However, continuing this trend is increasingly challenging, and the industry has predicted a slow-down in this trend over coming decades.}
\label{fig:density}
\end{figure}

Since the mid-2000s, however, trends in processor scaling have shifted because of the end of a phenomenon referred to as “Dennard scaling.” During the period of Dennard scaling, processors improved exponentially, not only in density, but also in clock speed. As fundamental physical constraints ended this trend, the density of digital electronics has continued to improve, but improvements to clock speed have been more modest. Furthermore, driven by technological and economic conditions, continuing improvements in density ("Moore's law" scaling) has become increasingly challenging, with slowdowns from historical trends predicted (Figure \ref{fig:density}) \cite{more_moore_team_international_nodate}. Even with density improvements, some circuits (particularly memories) have ceased displaying benefits from scaling \cite{schor_iedm_2022}. 

These shifts have motivated a growth in the diversity of commercial computing technologies, where processing performance can be gained not only from continued scaling of circuits but also from specialized instructions, representations, and circuit topologies \cite{hennessy_new_2019}. For instance, high-performance central processing units (CPUs) are multicore and require parallel programming techniques to fully utilize. Similarly, the main application of graphics processing units (GPUs) shifted to general-purpose parallel computing and now dominates the majority of available capability in high-performance computing systems \cite{top500_authors_top500_2023}. In particular, the growth of machine learning (ML) techniques has boosted the popularity of GPUs and driven the growth of accelerators specialized for ML tasks \cite{reuther_ai_2021, reuther_survey_2020}. Even given this variety of commercial solutions, engineers may find available products do not meet their requirements and must utilize a more customized solution for a given task. Previously, the design and fabrication of custom chips was a highly expensive undertaking, but advances in open-source instruction set architectures, high-level synthesis design tools, and process development kits have begun to lower the barrier of entry to creating custom processing solutions implemented via application-specific integrated circuits (ASICs) or field-programmable gate arrays (FPGAs) \cite{greengard_will_2020,bachrach_chisel_2012,izraelevitz_reusability_2017,fahim_hls4ml_2021,kahng_openroad_nodate,zhang_open-source_2022}.

Each of these components remains fundamentally based on the same technology: digital electronics. Therefore, we project each of these components’ capabilities for scaling similarly to the gradually slowing improvements in digital electronics. To predict the demand of future electronics, we assume that circuit density will continue to increase while maintaining an approximately constant per-element operating speed and total power dissipation level. CPUs, GPUs, and accelerators will provide more computing throughput (operations, `ops') at approximately constant speeds and total power, FPGAs will provide more programmable logic in the same area, and ASICs will implement the same function in smaller areas. 

In concert with processing, communication technologies play a crucial role in enabling computational systems. These technologies enable the transport of information between components in a computing system and include both short-range buses (such as Peripheral Component Interconnect Express - PCIe) and long-haul links (such as ethernet or InfiniBand). Edholm’s law of bandwidth describes the scaling of wired and wireless communication which, similarly to Moore’s law, has demonstrated exponential growth in the capacity of communication channels \cite{cherry_edholms_2004}. Fast and efficient communication is needed for computing systems across multiple length scales, as demand for bandwidths between heterogeneous computing modules and memories has continued to increase, particularly in the realm of machine learning (ML). This has driven the development of new standards and technologies for communication on short scales. The Universal Chiplet Interconnect Express (UCIe) is one such standard. It aims to provide a method for efficiently creating very dense, high-bandwidth links between chips such as CPUs, GPUs, and memory modules, enabling higher-performance systems in a single package instead on a larger, less-efficient circuit board \cite{biswas_universal_2021}. Additionally, several novel communication technologies incorporate advances in areas such as optoelectronics and integrated photonic circuits to provide optical links specialized for short chip-to-chip links with dense integration \cite{pezeshki_high_2021, wade_teraphy_2020}. These technologies could enable future computing systems to integrate multiple types of processing and storage systems more closely and with higher bandwidths and efficiencies. 

The design and integration of novel technologies into a processing system require significant engineering effort. To evaluate the benefits and trade-offs offered by novel technologies, it is crucial to measure their ultimate impact on system performance in comparison with employing standard technologies. High-energy physics (HEP) is one data-driven discipline which has applied computer-based analysis and experimental control for decades. To demonstrate the impacts which novel technologies may have on future experiments, we utilize HEP data acquisition systems (DAQs) as a case study.

\subsection{Data \& Computing in High-Energy Physics}

\begin{figure}[htbp]
  \centering
  \includegraphics[width=\linewidth]{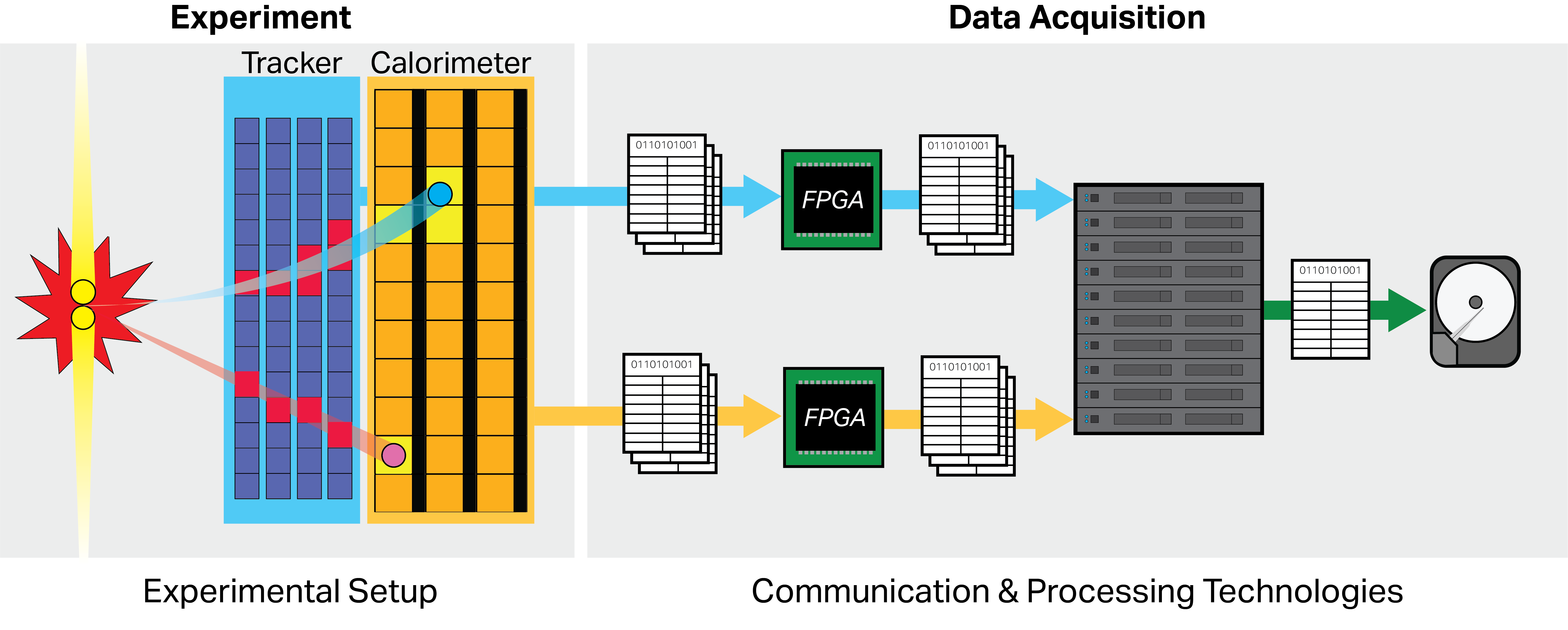}
  \caption{The amount of data that an experiment produces must be matched with a downstream computational system that meets or exceeds its needs. In the case of high-energy particle colliders, experimental variables such as luminosity and the resolution of detector systems influences the overall amount of data produced. Systems that analyze and classify the data require communication and processing systems. The power, area, and scalability of these systems are influenced by the specific technology being deployed.}
  \label{fig:pileup}
\end{figure}

The identification of fundamental physical particles and interaction mechanisms is a long-standing scientific goal, addressed over the previous century by refining theoretical models and collecting evidence to support or refute these theories. HEP has contributed to this process by producing and measuring fundamental particles such as the top quark and Higgs boson \cite{abachi_observation_1995, della_negra_journey_2012}. To accomplish this, bunches of particles are accelerated in well-defined "beams" to high energies. These beams then cross at precise locations, called "interaction points." At each crossing, individual particles which collide inelastically transform their energy into the creation of new particles that radiate outward \cite{wilson_introduction_2001}. These products are measured by detectors that surround the interaction point and measure properties such as electromagnetic traces and energy deposits \cite{cms_collaboration_cms_2008, atlas_collaboration_reconstruction_nodate}. These measurements allow for the identification of individual particles, their energies, and trajectories. Determining these parameters with high confidence enables a post hoc analysis to "reconstruct" the full interaction that led to a given set of products, identifying intermediate stages that are not directly measured by the detector (Figure \ref{fig:pileup}) \cite{cms_collaboration_particle-flow_2017}. 

However, the various physical interactions that yield observable particles do not all occur with equal probability; each process has a “cross-section” that quantifies the chance it will occur \cite{martin_parton_2009}. In order to improve the characterization of rare and potentially unknown processes, particle colliders have evolved to produce higher-energy collisions at a greater rate. The world’s current largest particle collider, the Large Hadron Collider (LHC), collides bunches of protons at a rate of 40 million per second (40 MHz) with a center-of-mass energy of 13.6 TeV. Within two crossing groups of protons, approximately 60 are expected to collide inelastically; this quantity is referred to as the "pile-up." Combining these factors and geometry of the beams yields the "instantaneous luminosity" of the system, which was measured in the most recent LHC run (Run-3) as up to $22.6 \cdot 10^{33}$ $cm^{-2} \cdot s^{-1}$ (Figure 2) \cite{herr_concept_2006, atlas_collaboration_peak_nodate}.

The Compact Muon Solenoid (CMS) experiment is one of two general-purpose detectors installed at the LHC. Under current conditions, the detector records approximately 2.0 MB of data each time two bunches of particles cross at the interaction point \cite{cms_collaboration_phase-2_2021}. These samples, termed “events” in HEP, are produced at the bunch crossing rate of 40 MHz, producing a combined output rate of approximately 80 terabytes per second, producing a volume of data which cannot sustainably be stored over months of experimental runs.

However, counterbalancing this torrent of data is the scarcity of results that are believed to contain novel data relevant to current physics experiments. In other words, evidence suggests that the majority of samples only provide measurements of particles and mechanisms that are already well characterized, yielding no novel scientific information. Therefore, in order to maintain a sustainable output, the CMS detector’s current (Phase-1) processing workflow (referred to as the Data Acquisition System or DAQ) is configured to ultimately save only 1 out of every 40,000 events, selecting for later analysis  only those with specific signatures \cite{das_overview_2022,cms_collaboration_cms_2017}. Currently, most DAQ systems utilize multiple processing stages to carry out this selection process in real time (Figure \ref{fig:pileup}). 

The CMS DAQ system utilizes both commercial and specialized electronic logic to carry out this processing. Novel technologies, including new ASICs, accelerators, and neuromorphic systems, could potentially be integrated into this system. To evaluate the impact such a component could have on the overall performance of the DAQ (and other real-time computing systems), a systematic evaluation model is needed. Applying our performance model, we estimate the impact of new technologies on this system in a case study below.

\section{The SystemFlow Model}
\subsection{Overview}
\begin{figure}[htbp]
  \centering
  \includegraphics[width=1.0\linewidth]{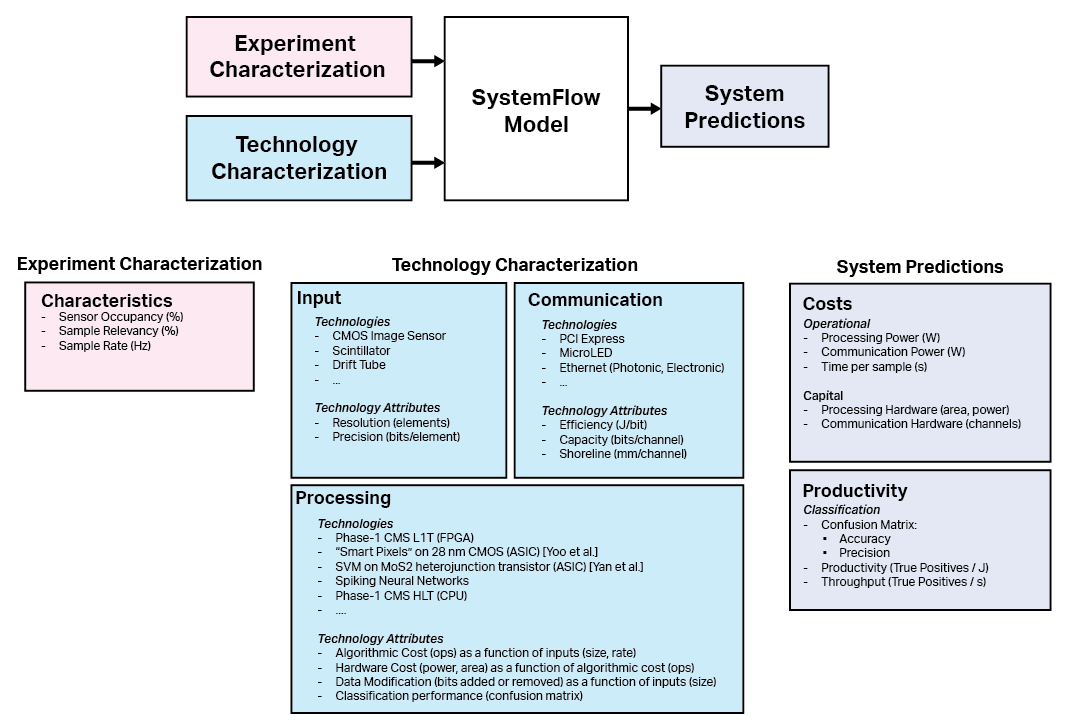}
  \caption{In a SystemFlow model, components of a computing system are classified into inputs, processing, outputs, and communication. Each category utilizes metrics and functions to capture how it transforms information and the requirements needed to do so. Each alternative technology for sensing, communication, and processing will have a unique set of attributes. By simulating the flow of messages containing information through this system, the attributes of each component interact to estimate component and system-level costs and productivity metrics (such as processing power, communication channels, and number of true positive samples produced per second).}
  \label{fig:systemflow}
\end{figure}

Most modern scientific experiments produce digital data which must be analyzed and stored by a computing system. In order to avoid losing these valuable data, a computing system must be matched to meet the requirements of an experiment. We utilize several key descriptors to capture both the needs of a scientific experiment and the attributes of technologies which may be deployed to meet them. These descriptors are summarized in Figure \ref{fig:systemflow}.

\subsection{Experimental Characterization}
A scientific experiment which produces digital data has several basic properties which predict the basic needs a computing system must meet. We predict these needs through the metrics of sensor occupancy (the proportion of sensing elements producing a non-zero result), the maximum number of experimental samples produced per second (Hz), and the proportion of samples which are relevant to scientific goals (\%). 

\subsection{Technological Characterization}
A computing system's components may be partitioned into four major categories: input, communication, processing, and output. Inputs collect information from the environment. Communication technologies enable the transport of information between different locations. Processing technologies utilize algorithms to transform information. Lastly, outputs present information to an external system or the environment. 

The arrangement of these components into a computing system may be represented as a directed graph, where inputs, outputs, and processing are represented by nodes and are connected via communication links (edges). In the case of data analysis systems for experiments, these graphs may take the form of trees, where information flows "upwards" from input sensors to intermediate processing stages, eventually coalescing at an output. The requirements and performance of the system depends on the interaction of each part as information flows up from the sensors towards storage, implementing a processing pipeline.

We model the propagation of information through a processing pipeline. This is accomplished by identifying the amount of information produced by each node and propagating it up the tree. Information is contained in discrete "messages," which originate at input nodes with a given frequency. Communication links transport these messages to processing stages. These stages implement algorithms which can change the amount of information in the message - increasing it by providing new analyses, or reducing it by applying compression or removing irrelevant data. Additionally, information from analysis may be used to make a decision, such as to discard a message. The propagation and modification of messages is simulated until each one is either discarded or reaches the output node. As this model describes the flow of messages through a computing system, we term this approach the "SystemFlow" model.

\subsubsection{Inputs}
We define an input as a node which records information from the environment with a set number of elements (resolution) and precision. Examples include a camera recording an image from millions of pixels with 8-bit color, or a spectrometer recording 1,024 frequencies with 16-bit precision. The constants defined by the experimental characterization and input design determine the size of messages ($bits$), proportion of messages relevant to the experiment ($\%$), and the rate at which they are introduced into the system ($Hz$). These parameters are determined empirically (experimental environment) or by the system designer (sensor design). 

\subsubsection{Communication}
Communication links are responsible for transporting information between different logical units in a computing system. These may take many different forms, from buses implemented via metal traces in a silicon wafer to long-range photonic links. Each is characterized by several parameters, chiefly latency ($s$), bandwidth ($bits/(second \cdot channel)$), efficiency ($J/bit$), and area requirements or "shoreline" ($mm/channel$). These are determined empirically for each available technology. 

\subsubsection{Processing} 
Processors implement algorithms chosen by system designers to carry out analyses relevant to a given message. Examples include searching for peaks in frequency spectra, object recognition, classification, and more. In hardware, each algorithm is realized by discrete processing steps which are taken for it to execute (ops). The number of ops may vary with the size of an incoming message and the algorithm; in the best case, ops remain constant with the size of an input. More often, they will scale linearly, polynomically, or even more aggressively. Furthermore, each algorithm may require greater or fewer ops for messages of identical size based on its contents (i.e. more `complex' data incurs more branching and processing steps). We avoid this latter complexity by assuming that the number of messages passing through the system is large, and an average processing time per message of a given size suffices to estimate a functional relationship between the two metrics ($bits$ and $ops$). Empirically or theoretically capturing the functional complexity for an algorithm on a processing technology translates informational requirements into hardware requirements; capturing this relationship is necessary to estimate how changing experimental conditions will induce new requirements for processing hardware. Knock-on requirements for each op, such as silicon area, power, and cost, can then be estimated for each hardware system. Functional scaling of ops and knock-on requirements varies for different technologies (e.g. CPU, GPU, ASIC) and may be modeled empirically and/or theoretically. 

After it has executed the necessary algorithm, a processing node will pass a message containing a result to the next node up the tree via a communication link. The size of this message will vary based on the processing stage; analyses may add information (such as high-level objects recognized) or remove information (transforming a full vector of wavelengths into a small list of peaks). This is measured in bits and may remain constant or vary with the size of the input. 

Finally, one subset of algorithms induce an additional consideration: classification algorithms can determine whether an incoming message is relevant to an experiment. When irrelevant, a message is dropped at the given processing node and does not propagate up the tree towards the output. To capture the performance of this classification process, the distribution of scores produced by the classifier analyzing relevant and irrelevant populations is modeled. Capturing this relationship is necessary to estimate how modifying a classifier to vary its sensitivity threshold will impact the performance of the overall system. These score distributions may be estimated empirically or parametrically, and produce a confusion matrix for each selected threshold. 

\subsubsection{Output}
The messages which propagate up the processing tree are captured as the system's output. This supplies the rates which must be sustained by the storage system ($bits/s$), as well as the number of relevant and irrelevant messages (true and false positives) which have propagated up the processing tree. These statistics may be used to estimate the overall productivity of the pipeline by calculating accuracy, precision, and these statistics normalized by overall cost (operational and capital expenditure). 

\subsection{Model Outputs}

To summarize, the propagation of messages through the system models the knock-on effects of multiple processing stages, each with unique costs and performance characteristics. These properties may be investigated in the model at each stage, or across the system as a whole. For instance, the power of a single processing stage or number of channels in one communication link may be inspected, as well as the power of the system as a whole or its overall performance at separating significant and insignificant samples. 

Furthermore, these estimates scale to different system-level design parameters, such as the technology utilized at each stage and varying experimental conditions. This enables system architects to not only model a system in its current configuration, but estimate how its requirements and performance will vary as system technologies or experimental conditions are altered.

\section{Case Study: The CMS DAQ}
\subsection{Current (Phase-1) DAQ}
The objective of the CMS DAQ is to maximize the capture of novel information for physics experiments (such as characterizing the Higgs boson and searching for evidence of supersymmetric particles). These data may be rare; for instance, models predict that currently the LHC produces a Higgs boson only once every few seconds \cite{stirling_parton_nodate}. With samples being produced at a rate of 40 MHz, this implies the selectivity of the DAQ must exceed parts per billion. This leads to strict requirements for algorithms which can detect relevant features from sensor data with high confidence and speed. The CMS DAQ utilizes two major processing stages to achieve this high selectivity: the Level-1 Trigger (L1T) and High-Level Trigger (HLT). 

In its current, or "Phase-1" DAQ configuration, the L1T utilizes information from the CMS’s muon and calorimetry systems to infer the presence of particles with high momenta (such as muons, tau leptons, electrons, and photons), as well as groups of particles with high energy (such as hadronic jets). These markers indicate that a rare, high-energy interaction may have taken place. Algorithms to examine detector data for these features are implemented in FPGA hardware located near the detector but in a radiation-safe zone. These results are used to down-select samples produced by the detector, with algorithms calibrated to pass, on average, one in every 400 samples to the next stage of analysis. 

The HLT utilizes more complex algorithms and all available data from CMS detector systems to “reconstruct” the event that led to the patterns recorded by the detector. This is a much more energy-intensive process than the simple feature detection used in the L1T; in the Phase-1 HLT, software implementing the reconstruction algorithms is run on commodity hardware (CPUs) in a conventional data center. This process further down-selects samples by a ratio of 1 to 100, leading to a final sample output rate of 1 $kHz$. The Phase-1 configuration is utilized as the "baseline" against which alternate systems are compared.

\subsection{Future HL-LHC Experiments}
Currently, the LHC is undergoing its final experimental run (Run-3) before upgrades are made to its accelerator systems. The objective of these upgrades is to increase the precision with which bunches are collided in the detector, leading to more high-energy collisions per bunch crossing. This will increase the overall luminosity of the LHC, leading to its `high-luminosity' (HL-LHC) configuration. In subsequent HL-LHC runs in 2028 (Run-4) and 2032 (Run-5), the stated goal is to achieve pile-ups of 140 and 200, respectively \cite{cms_collaboration_phase-2_2021}. 

Under the HL-LHC configuration, the CMS detector will record more data per sample: an increase in pile-up leads to more particles yielded from each collision, causing the tracker to register more charged tracks and the calorimeter to record more energy deposits. This will lead to an increased size of 8.0 MB per sample and total detector data rate of 320 TB/s during Run-5 \cite{cms_collaboration_phase-2_2021}. Furthermore, increasing the number of particles also leads to greater complexity in analysis. Simulations show that HLT analysis of Run-5 events will take approximately 16 times longer to analyze than Run-3 events with current algorithms and hardware \cite{cms_collaboration_phase-2_2021}. Finally, the number of events passed from the L1T to HLT is planned to increase 7.5 times, increasing the L1T output rate from an average of 100 $kHz$ to 750 $kHz$. This motivation behind this change is to utilize the higher selectivity of the HLT to retain events which might otherwise be rejected by the L1T and to provide more experimental data. 

These factors will increase the computational load which the DAQ must sustain; to quantify this increase, we adapt the current (Phase-1) CMS DAQ into a SystemFlow model. The parameters of this model are then shifted to simulate the conditions under the final planned HL-LHC configuration, Run-5. 

\subsection{SystemFlow DAQ Model}
\begin{figure}[htbp]
  \centering
  \includegraphics[width=0.6\linewidth]{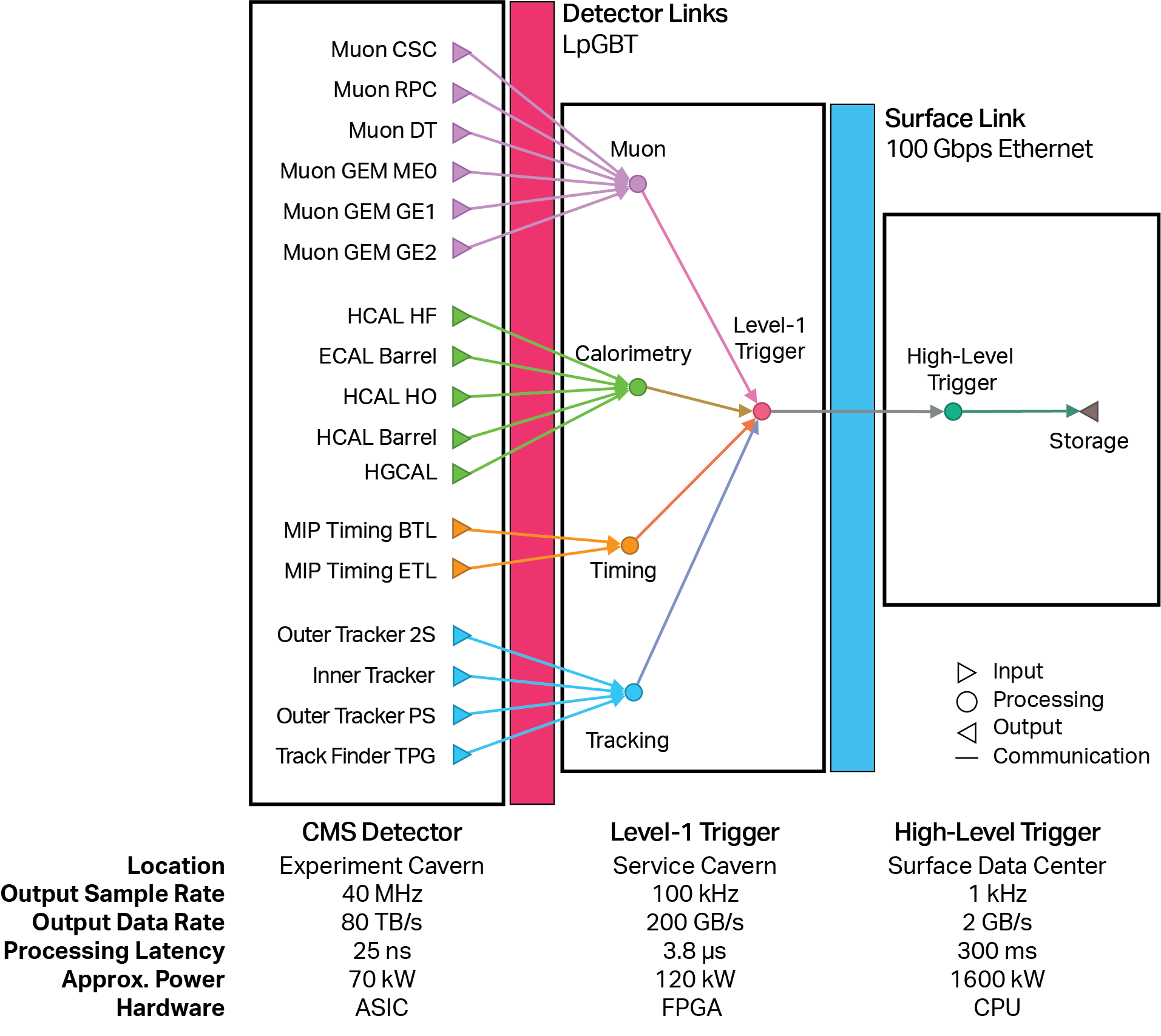}
  \caption{An illustration of the CMS DAQ system adapted into a SystemFlow model. Samples originate at sensors providing input to the computing system (left). These samples are transported via communication links to processing nodes (middle), where they may change in size and/or be discarded. These changes propagate downstream to further processing nodes, and eventually, the output (right). The metrics necessary to characterize each node and edge in the SystemFlow model are taken from public documentation.}
  \label{fig:cms_systemflow}
\end{figure}

The CMS DAQ is adapted into a SystemFlow model by utilizing a breakdown of CMS detector sub-systems to identify all input nodes and the amount of data they produce. Processing nodes are created for each detector sub-system, the L1T, and HLT. The HLT provides the output of the system which is archived \cite{cms_collaboration_cms_2008, cms_collaboration_phase-2_2017, cms_collaboration_phase-2_2020, cms_collaboration_phase-2_2021}. This SystemFlow model of the CMS DAQ is shown in Figure \ref{fig:cms_systemflow}.

\subsubsection{Detectors}
All detector subsystems within the CMS are included as input nodes in the SystemFlow model. The effect of pile-up on amount of incoming data is included by scaling the bits recorded by each sensor by pile-up according to CMS estimates \cite{cms_collaboration_phase-2_2021}. A new detector system, the high-granularity calorimeter (HGCAL) will be installed in the CMS after Run-3 and incorporated into the detector for Runs 4 and 5. The significant amounts of data data produced by this subsystem are included in the estimates for Runs 4 and 5.

\subsubsection{L1T}
Data is transported from sensors to the L1T by a custom radiation-hardened link, the low-power GigaBit Transceiver (LpGBT). Public data places the efficiency of this link at 22 pJ/bit \cite{cms_collaboration_phase-2_2020}. Algorithms used within the L1T identify features such as high-energy particles and "jets" \cite{das_overview_2022}. These relatively simple algorithms search for clusters and peaks, and we estimate that the resources required for these algorithms will scale approximately linearly with the amount of incoming data from sensors. The performance of these algorithms at classifying the incoming samples as "relevant" or "irrelevant" is modeled in detail on the performance of identifying each high-energy feature. For details, see \nameref{sec:methods_classifier}. Power efficiency of the L1T is scaled to its current needs, approximately 120 $kW$ \cite{cms_collaboration_phase-2_2020}. Currently, the L1T only retrieves tracking data after identifying a feature of interest from other data, but this feedback behavior is not included in the current model. This produces no significant difference in the metrics predicted by the SystemFlow model, impacting only communication costs from the detector to the trigger which are less than 1\% of overall system usage. 

\subsubsection{HLT}
Data accepted by the L1T is transmitted to the HLT datacenter by standard 100 gigabit ethernet over fiber; commercially available, medium-link transceivers require approximately 25 pJ/bit \cite{noauthor_commercial_nodate}. Complex algorithms are utilized within the HLT to identify particle tracks, associate them with calorimeter hits, identify intermediate particles produced by the collision, and more. The scaling of these algorithms with increased incoming data is modeled empirically by collating data from CMS and ATLAS experiments estimating the run-time of HLT reconstructions under increasing pile-up \cite{cms_collaboration_phase-2_2021, atlas_collaboration_reconstruction_nodate}. Power efficiency of the HLT is scaled to its current consumption, approximately 1.6 $MW$ \cite{cms_collaboration_phase-2_2021}. The HLT is much more complex than the L1T, with hundreds of separate "paths" allowing an event to be selected for offline storage. The overall performance of the HLT as a classifier is modeled by utilizing public results detailing the performance of 6 different trigger paths used for different experiments within the CMS (e.g. Higgs, Supersymmetry, exotica - details in \ref{sec:methods_hlt}). 

\section{Results}
\subsection{Phase-1 System}
Firstly, we examine characteristics of the current (Phase-1) CMS DAQ as if it were scaled to meet the desired experimental goals for Run-5: a collider pile-up of 200, and decreasing the rejection ratio of the L1T from 400:1 to 53:1 (increasing the effective Level-1 "trigger rate" from 100 to 750 $kHz$ and overall output of the DAQ to 7.5 $kHz$). Hardware and algorithms within the system are held constant, but hardware during Run-5 (2032) is modeled as being 6.5 times more energy-efficient than today (2024) based on a corresponding increase in the areal density of integrated circuits operating at constant power (see \nameref{sec:computing_trends}). DAQ productivity is measured in the expected number of relevant samples produced per second normalized by the power taken to acquire them $(1/J)$. This is produced from each SystemFlow model by multiplying the total output rate by its system-level F1 score ($1/s$), normalized by power used across the entire pipeline ($J/s$). 

\[productivity = \frac{output \, rate}{power} \cdot \frac{2 \cdot precision \cdot recall}{precision + recall} \]

\begin{figure}[htbp]
  \centering
  \includegraphics[width=0.6\linewidth]{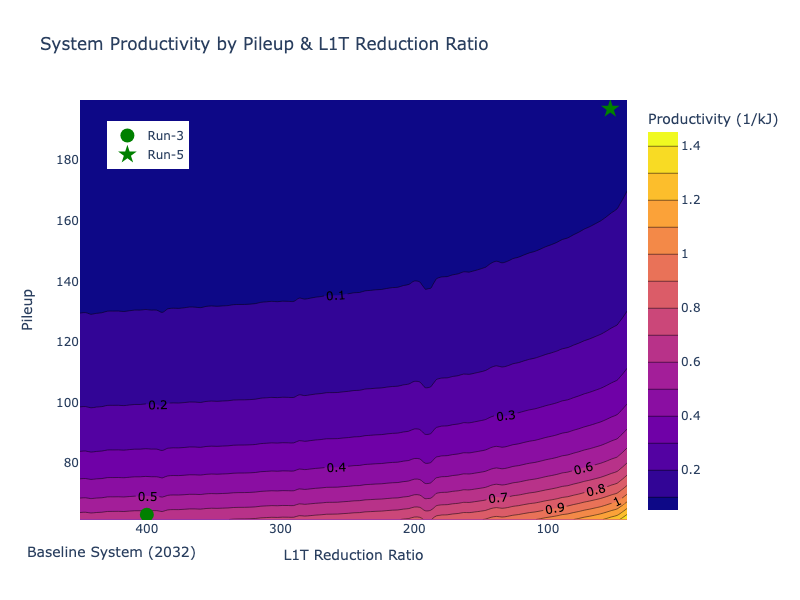}
  \caption{Increasing the pileup of collisions within the detector and increasing the number of samples passed from the L1T to the HLT significantly changes the productivity of the overall DAQ system. As conditions shift from Run-3 (bottom left) to Run-5 (top right), net productivity first increases by reducing the L1T's rejection ratio. However, the energy cost of processing more samples within the HLT as pile-up increases causes productivity to drop steeply.}
  \label{fig:phase-1}
\end{figure}

Meeting the Run-5 experimental goals will have a large impact on the DAQ. Increasing pile-up from 60 to 200 greatly lengthens the amount of runtime needed to process a sample within the HLT, due to the increased complexity of the underlying data - in particular, processes such as reconstructing particle tracks scales poorly with increasing pile-up \cite{trackml_accuracy, cms_collaboration_phase-2_2020}. Furthermore, to retain increased amounts of relevant information, 7.5 times more samples will be sent from the L1T to the HLT. Combining these changes causes overall DAQ power requirements to increase significantly, from 0.32 $MW$ to 52 $MW$, with the increase almost entirely stemming from CPU-based HLT processing (Table \nameref{tab:baseline}). Productivity experiences a net drop from 0.80 to 0.065 $1/kJ$ due to the greatly increased power requirements for a moderately increased rate of relevant samples retrieved by the system (Figure \ref{fig:phase-1}).

\begin{table}
\centering
\caption{SystemFlow Estimates of Phase-1 DAQ Productivity by Pile-Up and L1T Reduction Ratio}
\smallskip
\label{tab:baseline}
\begin{tabular}{cc|cllcc}
      Pile-up&\makecell{L1T \\Reduction Ratio}&\makecell{DAQ Power\\(MW)}&\makecell{Precision\\(\%)}&\makecell{Recall\\(\%)}&\makecell{F1 Score\\(\%)}&\makecell{Productivity\\(1/kJ)}\\ \hline
      60&400:1&
 0.32&   28.&28.&28.&0.86\\
 200& 400:1& 7.0&   23.&24.&24.&0.034\\
 200& 53:1& 52.&   40.&40.&45.&0.060\\\end{tabular}
 \end{table}

Scaling the baseline DAQ system to the needs of Run-5 is thus undesirable, representing approximately a 20 times increase in power expenditure compared to the Phase-1 DAQ during Run-3. This motivates the changes to the system whose impacts we investigate next: utilizing GPU hardware within the HLT, incorporating tracking into the L1T, and utilizing smart sensing elements within the detector.

\subsection{Next-Generation Systems}
We quantify the impact of three major changes proposed to improve the CMS DAQ as it scales to meet Run-5 conditions:  incorporating tracking information into the L1T, utilizing GPU-based hardware in the HLT, and utilizing detector-integrated data filtering in the CMS Inner Tracker ("smart" sensing). The impact of these individual changes on the overall DAQ system is modeled by introducing corresponding changes into the SystemFlow model.

\subsubsection{L1 Tracks}
Tracking information is not utilized in the Phase-1 L1T, which makes certain analyses – such as distinguishing charged and non-charged particles (e.g. electrons and photons) – infeasible. Future upgrades to the L1T system focus on incorporating tracking information and making other algorithms available to improve the selectivity and capabilities of the L1T system \cite{cms_collaboration_phase-2_2020}. We explore the impact which improved L1T classification skill has on the DAQ by increasing the separation between scores produced by positive and negative samples in a model of the Phase-1 L1T classifier (see \nameref{sec:methods_classifier} for details).

\begin{figure}[htbp]
  \centering
  \includegraphics[width=0.6\linewidth]{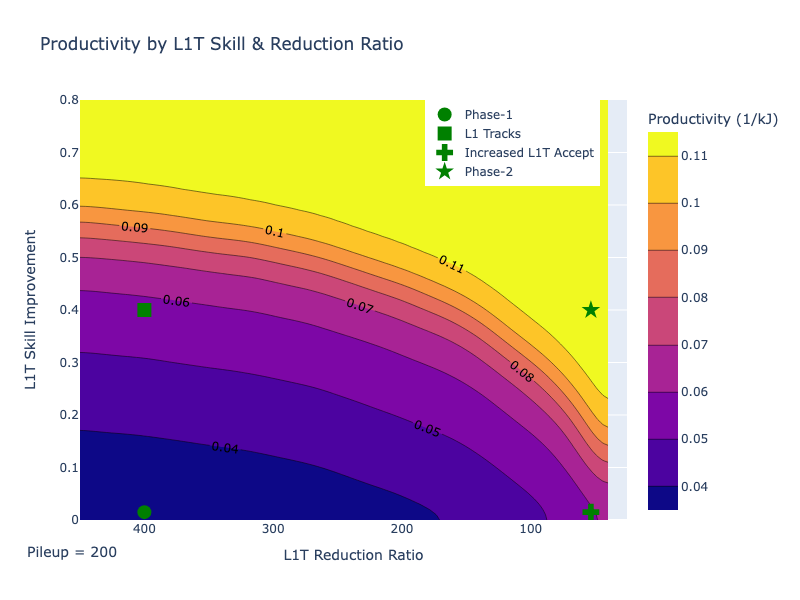}
  \caption{Productivity of a DAQ system operating on an experiment with an average pile-up of 200 is plotted as the reduction ratio of the L1T is reduced and its skill is increased. Introducing higher skill into the L1T is highly beneficial for overall system productivity, yielding greater gains than simply passing more samples to the HLT.}
  \label{fig:l1t}
\end{figure}

\begin{table}
    \centering
    \begin{tabular}{c|cccc} 
         System&  DAQ Power (MW)&  Precision (\%)&  Recall (\%)&  Productivity (1/kJ)\\ \hline
         Phase-1&  52.&  40&  40&  0.059\\ 
         GPU HLT&  26.&  38&  39&  0.11\\ 
         L1 Tracks&  52.&  100&  79&  0.10\\ 
         Smart Pixels&  41.&  40&  40&  0.074\\ 
         GPU and L1 (Phase-2)&  26.&  100&  79&  0.20\\ 
         Smart GPU&  41.&  100&  79&  0.13\\ 
         Smart L1&  20.&  38&  39&  0.14\\ 
 Smart Phase-2& 20.& 100& 79&0.26\\ 
    \end{tabular}
    \caption{System-level power consumption, classification performance, and overall productivity as various changes are implemented into the CMS DAQ. "GPU" corresponds to the utilization of GPU hardware to parallelize HLT algorithms for event selection. "L1 Tracks" corresponds to the introduction of tracking information in the L1T, improving the performance of its classifier. "Smart pixels" corresponds to the introduction of smart sensing elements within the inner tracker that remove unnecessary data from samples.}
    \label{tab:productivity}
\end{table}

Our model predicts that improving the performance of classification within the L1T can lead to significant improvements in overall system performance; improving the separation of scores between positive and negative samples within the L1T by 40\% can lead to the same improvement in overall system classification performance as increasing the number of samples sent from L1T to the HLT by 750\% (Figure \ref{fig:l1t}). As the L1T is responsible for discarding the majority of samples collected from the detector, any improvement to it significantly improves the entire DAQ.

\subsubsection{GPU HLT}
Many algorithms utilized within the event reconstruction carried out in the HLT can be effectively parallelized; this allows software to take advantage of hardware such as GPUs to increase throughput. Experiments using parallelized routines running on GPU hardware for CMS event analysis show an increase in the number of samples which can be processed per second by using a GPU (such as an approximate 50\% increase using an NVIDIA A100) \cite{cms_collaboration_phase-2_2021}. This change is captured in the SystemFlow model by adjusting the complexity function of the HLT processing node to match the throughput results of the parallelized algorithms. Simultaneously, the power for the node is modified to match the average dissipation of the NVIDIA SXM4 A100 accelerator (400 $W$). Incorporating this change into the DAQ system would roughly double system productivity to 0.13 $1/kJ$, as HLT consumes the vast majority of processing energy (Table \ref{tab:productivity}). Aggressive improvements to GPU hardware could further improve the energy efficiency of traditional HLT algorithms - however, this may be limited by development trends in GPUs focusing on limited-precision numerical representations adapted to AI systems' needs \cite{dally2023hardware}. 

\subsubsection{Smart Sensing}
Processing adapted to specific sensors has been proposed as a method to reduce the amount of data which must be transported from detector systems while simplifying subsequent processing. For instance, high-resolution pixel sensors with on-chip neural networks can be used to detect and reject data from low-momentum particles irrelevant to later analysis \cite{yoo_smart_nodate}. The impact of this change on the DAQ is modeled after this result by assuming neural network ASICs integrated into next-generation tracking modules ("Smart Pixels") could reduce the amount of data from an inner tracker pixel module by 54\% or more, dissipating 300 $\mu W$ to analyze 256 pixels. We assume this approach could scale effectively to the CMS Inner Tracker, which has two billion pixels and would require at least 2.3 kW to implement a classification algorithm co-located with detection elements.


We estimate that for the cost of dissipating several kilowatts of power in the detector, a “smart pixels” approach discarding low momentum particles constituting 54\% of tracking data could ultimately lead to a net reduction of 6 MW over the next-best DAQ system, greatly increasing overall system performance (Table \ref{tab:productivity}). This saving is enabled by leveraging a simple algorithm adapted to specialized hardware utilizing local information to avoid more complex analysis later in the pipeline. 

\subsection{Discussion}
Through our case study, we have demonstrated that depending solely upon Moore's Law scaling to meet the future demands of HEP experiments would require greatly increased capital and operational expenditures if changes to the processing system (DAQ) are not implemented. As the cost of computing becomes a larger issue in society, simply accepting these increased expenditures is not an acceptable solution to meet these future demands.

However, in anticipation of this challenge, several approaches are being developed. We modeled the impact of improving the DAQ's L1T, utilizing GPU-based processing in the HLT datacenter, and implementing "smart" detector elements in the front-end of the system. Each of these changes carries unique consequences; improvements to the L1T greatly improve the quality of results sent to the next stage of processing, the HLT, sending fewer false positives and discarding fewer false negatives with minimal impacts to system power. The HLT's efficiency can be improved through the use of parallelized computation carried out on GPU, roughly halving the amount of power utilized by the DAQ. Net system power can be further reduced by eliminating unneeded data from samples by using "smart" sensing strategies. 

Overall, including these three changes greatly boosts the overall productivity of the DAQ to 0.31 1/kJ (Table \ref{tab:productivity}; this exceeds our estimate of the current productivity of the Phase-1 HLT, 0.13 1/kJ. However, given the objective of the DAQ to not only process more complex samples created under the conditions of increased pile-up, but to produce more samples of interest to experimenters, this productivity must be further improved. To reach the goal of producing 7.5 thousand relevant samples per second without significant increases to power, the DAQ must reach 0.98 1/kJ. Achieving this further 3 times improvement may require more aggressive changes to the DAQ such as adapting approximate or limited-precision computations, as well as utilizing machine-learning assisted triggers. We suggest that in particular, further improvement of the Level-1 Trigger’s skill via ML-augmented methods could provide significant advantages over decreasing its rejection ratio. Even utilizing very small networks, ML methods can identify advanced features such as top quark jets \cite{Bogatskiy_Hoffman_Offermann_2023}. This could allow for advanced triggering of objects of interest within the L1T while utilizing small amounts of power and enabling execution at low latencies suitable for the L1T, bringing HLT-like capabilities earlier in the pipeline.

\section{Conclusion}
The amount of data collected by scientific experiments will continue to grow, as will the need to process these data. Simultaneously, the capabilities of computing workflows in the future will become increasingly dependent on the application and integration of specialized hardware, creating heterogeneous systems. However, the design and integration of specialized hardware can require significant investment. In order to justify the design of a heterogeneous system, the opportunities and trade-offs it offers over conventional approaches must be rigorously projected.

To address this need, we describe the “SystemFlow” framework, an approach which allows computational workflows to be modeled using high-level characteristics. This allows for rapid investigation of alternative system configurations using preliminary data and results. We apply this approach to characterize a high-energy physics experiment and project the impact of integrating GPU processing, improved early-stage classification, and front-end data reduction into this system. We show that these integrations could provide major reductions in the power required for the system to operate (approximately 60\%), but that more advances will be required in order for the system to match or surpass the productivity of earlier experiments running under less challenging experimental conditions. 

\appendix

\section{Research Methods}
Code is available online at \href{https://github.com/wilkieolin/system\_flow}{https://github.com/wilkieolin/system\_flow}. 

\subsection{Classifier Model}
\label{sec:methods_classifier}
Messages which represent data which should retained by the data acquisition system (DAQ) and messages which should be discarded represent two distinct populations. Each processing node in the system may implement a classifier which seeks to determine which population a message is a member of, either tagging it to be retained and sent to the next node in the workflow graph or discarded. 

 Various strategies may be used to implement a classifier, seeking to trade-off accuracy for speed or processing requirements. To avoid the need to re-implement specific details of each system, each is interpreted as fundamentally being capable of producing a “score” for each incoming message. Ideally, messages to be discarded will produce a score of zero, producing a distribution of statistics which represent the “false” samples. Ideally, “true” messages to be retained produce score values which are high and easily distinguishable from scores in the false distribution. The cumulative distribution function (CDF) of each population’s scores may be calculated from parametric or empirical distributions:
\[
    CDF_{\textbf{Z}}(z) = P(\textbf{Z} \leq z)
\]

Where $\textbf{Z}$ represents scores produced by the classifier on a given population and $z$ represents an individual score.  

Each classifier must independently determine whether to pass or reject a message based on the score it produces, requiring a “threshold” to be set. Messages producing scores above this threshold will be accepted, and those falling below it, rejected. The value of this threshold will affect both the amount of data which is sent by one node in the workflow to the next node and the classifier’s performance. The independent variable, determined by the system-level requirements, is the amount of data which the classifier sends to the next processing node; classification performance responds to this parameter. 

To determine the performance of each classifier, the distribution of scores for the false and true message populations are weighted by the number of incoming true and false messages:

\[
    CDF_{node}(z) = \frac{n_{true} \cdot CDF_{true}(z) + n_{false} \cdot CDF_{false}(z)}{n_{true} + n_{false}}
\]

The threshold for classification ($z_{T}$) will be set at a score which rejects the desired number of samples:

\[
    z_{T} = argmin \left(\left|\frac{n_{rejected}}{n_{inputs}} - CDF_{node}(z)\right|\right)
\]

This determines the proportion of samples from the underlying distribution which will be retained or rejected:

\[
    TN = n_{false} \cdot CDF_{false}(z_{T})
\]
\[
    FP = n_{false} \cdot \left( 1 - CDF_{false}(z_{T}) \right)
\]
\[
    TP = n_{true} \cdot \left( 1 - CDF_{true}(z_{T}) \right)
\]
\[
    FN = n_{true} \cdot CDF_{true}(z_{T})
\]

The rate of true and false negatives and positives may be used to construct a contingency matrix for the classifier at a given threshold. Messages classified as false are discarded, and messages classified as true are passed to the next classifier. 

To obtain the contingency matrix for the system as a whole, true negatives and false negatives are summed across the entire system, and positives are only produced by the final output where messages are saved to disk. 

\subsection{Level-1 Trigger}
\label{sec:methods_l1t}
Accurately capturing a distribution of scores produced by classifiers analyzing messages is key to implementing a model which can faithfully produce system-level performance. For this reason, the Level-1 Trigger (L1T)’s performance is captured in detail to produce these scores.

The Phase-1 L1T centers on detecting the existence of certain features: electrons, photons, muons, tau leptons, and hadronic jets with momenta or energy sums above given thresholds (30 GeV for electrons and photons, 22 GeV for muons, 38 GeV for tau leptons, and a total of 320 GeV for hadronic jets). The L1T does not directly produce scores, but accepts any event with one or more of these signatures. To instead produce scores, we examine the efficiency curves of each of these object detections ("trigger paths") and interpret them as probability that a particle with a given momentum will be correctly detected. To produce a distribution of near-threshold events, an exponential distribution of particle momenta is fitted to reproduce the trigger rates produced by the L1T during the LHC Run-2 (2018):

\[
rate_{object}(\lambda) = \int_{0}^{\infty} efficiency_{object}(p) \cdot PDF_{object}(\lambda, p) \cdot dp
\]
\[
PDF_{object}(\lambda, p) = \lambda \cdot exp(-1 \cdot \lambda \cdot x) \cdot (x > 0)
\]

\[
\lambda_{object} = argmin \left( || rate_{object}(\lambda) - rate_{object, empirical} || \right)
\]

To generate a model of scores produced by classifying objects within the HLT, particles are randomly generated from the distribution of object momenta fit to the triggers. Objects which fall below the selection threshold represent the null distribution (samples to be discarded), and objects with momenta above the threshold represent the positive distribution (samples to be retained). By arbitrarily generating many samples ($n = 50,000$), the distribution of null and positive scores for each L1T object detection trigger is generated. These null and positive distributions are summed to calculate overall L1T scores for each sample and used to produce the confusion matrix of the L1T at a given selection threshold.

\subsection{High-Level Trigger}
\label{sec:methods_hlt} 
The high-level trigger (HLT) contains hundreds of different features (the HLT "menu") which may lead to a sample being retained for offline storage and analysis. Modeling in full each segment of this menu is beyond the scope of this work, and the HLT is approximated utilizing results from the accepted Run-2 (2018) performance of the CMS HLT. Six representative triggering paths for different physics objects (B2G/Exotica, Higgs, SUSY, Muons, Tracking, and Tau) are selected, and the efficiency curves of each path fit to an exponential distribution of objects in the same manner as the L1T. However, each path is selected independently, and scores are not summed across each trigger but taken one-at-a-time to generate the distribution of null and positive scores applied to generate a confusion matrix for a given selection threshold.

\subsection{Quantifying the Cost of Classifier Errors}
Discarding a data sample which should have been retained (a false negative) carries a different energy cost than retaining a sample which should have been discarded (false positive). This difference of energy within the data processing system can be quantified using the SystemFlow model.

As messages pass through the system, they may be discarded at any processing node. The farther the message propagates through the system, the more energy is used to process and transmit it. The  total energy $TE$ utilized to reach any node $n$ in the computational graph $\mathcal{G}$ can be found by traversing the paths available to reach it:

\[
TE(\mathcal{G}, n) = E_{n} + \sum{TE(predecessor(\mathcal{G}, n))}
\]

The mean energy taken to reach a node is the average of the total energy taken by each of its predecessor nodes, weighted by the number of messages $N$ from each node:

\[
MTE(\mathcal{G}, n) = \frac{1}{\sum_{p}^{predecessors(\mathcal{G}, n)}{N_p}} \cdot \sum_{p}^{predecessors(\mathcal{G}, n)}{TE(\mathcal{G},p)}
\]

Using the energy taken to reach a given node, we may calculate the differential in power between errors of type I (false positives) and errors of type II (false negatives). These differentials are made in comparison correct outputs (true positives and true negatives).

In the case of a true positive, a true message reaches the output node (root node of the tree graph). The energy taken is thus the mean total energy (MTE) of the output node:

\[
    E_{TP} = MTE{(\mathcal{G}, root(\mathcal{G}))}
\]

True negatives may be found not only at the output node, but at any processing node. The amount of energy produced by a true negative result is thus variable, but the expected value over many messages may again be found. The MTE of each classifier node is weighted by the number of messages it discards, producing an average energy consumed by each true negative:

\[
    E_{TN} = \frac{1}{\sum_{c}{classifiers(\mathcal{G})}{N_{c}}} \cdot \sum_{c}{classifiers(\mathcal{G})}{N_{c} \cdot MTE(\mathcal{G}, c)}
\]

The cost of false (incorrect) decisions is estimated in comparison to true (correct) decisions. In the case of a false positive, the message consumes the cost of a true positive while providing no useful information. Its cost is thus the differential between the amount of energy it consumed during processing and the amount it should have consumed:

\[
    E_{FP} = E_{TP} - E_{TN}
\]

In the case of a false negative, a message consumes the expected energy of a true negative while valuable information is discarded. In this case, the data processing consumes less energy, but this is countered by the fact that the system must keep running for longer to provide true, useful samples. The cost of a false negative is thus estimated as the cost of processing a true negative, plus the estimated cost of acquiring a "replacement" true positive. This replacement comes with both the cost of processing a true positive and several false positives, multiplied by the ratio of true negatives to one true positive at the output node. 

\[
    E_{FN} = E_{TN} + \frac{TN(root(\mathcal{G}))}{TP(root(\mathcal{G}))} \cdot E_{TN} + E_{TP}
\]

\acknowledgments
This work was supported by DOE ASCR and BES Microelectronics Threadwork. This material is based upon work supported by the U.S. Department of Energy, Office of Science, under contract number DE-AC02-06CH11357.

The authors would like to thank the following contributors for many fruitful discussions: Jennet Dickinson, Nhan Tran, Jieun Yoo, Jinlong Zhang, Alexander A. Paramov, Kevin Brown, and Tupendra Oli.

\bibliographystyle{JHEP}
\bibliography{SystemFlow}

\end{document}